\documentstyle[epsf,floats,prb,aps]{revtex}
\begin{document}
\title{Semiempirical Hartree-Fock calculations for pure and Li-doped KTaO$_3$}
\author{R.~I.~Eglitis,\cite{*} A.~V.~Postnikov, and G.~Borstel}
\address{
Universit\"at Osnabr\"uck -- Fachbereich Physik,
D-49069 Osnabr\"uck, Germany}
\date{25 November 1996}
\twocolumn[\hsize\textwidth\columnwidth\hsize\csname@twocolumnfalse\endcsname
\maketitle
\begin{abstract}
In extension of our previous study of KNbO$_3$
by the semiempirical Hartree-Fock method
we present parameterization and total-energy results
for non-ferroelectric KTaO$_3$ as a pure crystal
(concentrating on the frozen phonon calculations) and that
with Li impurities. The magnitudes of off-center
Li displacements and the relaxation energies related
to re-orientation of Li are calculated and compared
with experimental estimates and earlier calculation
results. The spatial extent of lattice relaxation
around Li impurities and contributions from different
neighbors to the relaxation energy are discussed.
\end{abstract}
\draft
]

\section{Introduction}
\label{sec:intro}
Potassium tantalate is a so-called incipient ferroelectric which develops
a high value of the dielectric constant and considerable softening of
a zone-center transverse-optical (TO) phonon,
typical to other ferroelectric materials, but stays in
a paraelectric state even at the lowest temperatures. This paraelectric
state can be modified by doping, that may (in case of Nb substituting Ta)
bring the system to a true ferroelectric state, or (for doping with
a sufficiently small amount of Li or Na that enter the K sublattice)
gives rise to a more complicated state, that is often referred to as
orientational glass (see, e.g., Refs.~\onlinecite{hkl90,vg90} for a review).
As was discovered by Yacoby and Just\cite{yac74},
a Li ion substituting K in KTaO$_3$, because of its smaller
ionic radius, gets spontaneously displaced along one of the (six)
[100] directions, towards an octahedral interstitial.
This effect has been further studied in
Refs.~\onlinecite{bor80,yac81,vdkr83}.
An experimental estimate of the displacement magnitude
possible from $^7$Li NMR measurements
(based on the measured values of the local electric field gradient,
interpreted in terms of ionic displacement) is 0.86~{\AA}
according to Borsa {\it et al.}\cite{bor80}.
Another estimate cited by van der Klink {\it et al.}\cite{vdkr83},
in addition to the former one, is $\sim$1.2~{\AA}.
H\"ochli {\it et al.}\cite{hkl90} report the displacement
magnitude of $1.1\pm0.1$~{\AA}.
These values are therefore to some extent dependent on the
model in question that provides the fit to the data
directly measurable by NMR.

Van der Klink and Khanna\cite{vdkk84} calculated the energetics
of the [100] Li displacement in a polarizable point-charge model,
with Coulomb and polarization energy treated as those from point
charges (of nominal ionic values), with a selected set of
polarizabilities. The equilibrium displacement (1.35~{\AA})
agrees well with the NMR-based estimation of $\sim$1.2~{\AA}, but the model
is probably too crude for handling a delicate balance between
long-range electrostatic forces and short-range effects of
chemical bonding in an incipient ferroelectric system.
Particularly, it became well known since then that effective
charges in perovskites deviate considerably from nominal ionic
values, and the lattice relaxation for instance in the
KTaO$_3$:Li system gives a substantial contribution to
the energy lowering on a Li displacement.

Several studies on Li in KTaO$_3$ have been done with
the use of the shell model which is more sophisticated than
the models dealing with point charges or point dipoles.
Stachiotti and Migoni\cite{stac90} calculated equilibrium
displacements of Li and neighboring atoms making use
of the static lattice Green function method formulated
on top of the nonlinear shell model, with anisotropic
core-shell couplings of the oxygen ion.
The drawback of this approach is that the anisotropic
(fourth-order) interaction was allowed for the O--Ta bond only,
but otherwise the model was fully harmonic. This seems
to be hardly satisfactory for quantitative estimates,
taking into account the large magnitude of the Li displacement.
Moreover, there was an ambiguity in fitting the Li--O
potential of the Born-Mayer type, based either on a study of
lattice dynamics in LiKSO$_4$\cite{char87}, or on the study
of defect energies of Li$^+$ substituting Ba$^{2+}$
in BaTiO$_3$\cite{lew86}. The results of Ref.~\onlinecite{stac90},
obtained with these two sets of parameters,
differ considerably: the first set leads to a non-displaced position of Li,
whereas the second set provides the [100] displacement
by 1.44~{\AA} that seems to be too large.

Exner {\it et al.}\cite{ex:Li} performed another simulation
of doped KTaO$_3$ within the shell model, with the polarizability
of the oxygen ion treated as isotropic, but, on the other hand,
with anharmonic effects accounted for. Also the choice
of short-range interaction was different in Ref.~\onlinecite{ex:Li}
from that of Ref.~\onlinecite{stac90}. The equilibrium
off-center Li displacement was found to be $\sim$0.64~{\AA},
and the relaxation pattern of near neighbors to Li impurity
essentially different from that provided
by Staciotti {\it et al.}\cite{stac90}.

First-principles supercell calculations\cite{wil94} by the full-potential
linear muffin-tin orbitals method (FP-LMTO) made it possible
to study the total energy as function of off-center Li displacements
from its K-substituting position in KTaO$_3$. This calculation,
although not dependent on the choice of the particular interaction model
or the fitting of the potential, had nevertheless its technical
limitations. For one thing, the computational effort rises quite fast
with the supercell size. The largest supercell treated in
Ref.~\onlinecite{wil94} included $2\!\times\!2\!\times\!2$ primitive cells,
i.e. only 40 atoms. Another difficulty relates to the {\it muffin-tin}
geometry used in FP-LMTO calculations (see, e.g., Refs.~\onlinecite{msm1,sav}).
This assumes all atoms to be
circumscribed by non-overlapping spheres of reasonable size,
so that the potential and charge density are expanded in
spherical harmonics inside such spheres, and the interstitial region
is treated in some different way. The large magnitude of the Li off-center
displacement and, as a result, of lattice relaxation makes the
universal non-overlapping packing for all geometries possible
only with quite small spheres, that is disadvantageous for the
numerical accuracy.

The method of our choice in the present study is therefore
a tight-binding scheme with the basis of atom-centered orbitals,
that makes no use of {\it muffin-tin} geometry whatsoever.
At the same time, we want the scheme to be not much computationally
demanding, in order to keep the calculations for large supercells
feasible. What in some aspects closes the gap between empirical
models (applicable to very large systems) and accurate {\it ab initio}
schemes (that become very computationally demanding unless applied
to supercells of quite modest size) are semiempirical methods
which are able to produce reliable quantitative predictions,
based on a limited number of basic (system-dependent) tunable
parameters. We use the Intermediate Neglect
of the Differential Overlap (INDO) method\cite{popl,indo1,indo2},
which has been applied very successfully for the study of defects,
both in the bulk and on the surface,
in many oxide materials\cite{indo1,indo2,indo3,zro2,skc94,coru2,tio2},
as well as semiconductors\cite{gap94,stki96}.
We recently applied this method to the study of KNbO$_3$\cite{indo}.
The method is essentially a simplified implementation
of the Hartree-Fock formalism, with its obvious drawback of
underestimating the correlation effects. The latter seems however
to be of little importance when one studies an ionic insulator
as KTaO$_3$ and not, say, a metallic system. Moreover, the
appropriate choice of parameters makes it possible to obtain
reasonable trends in the total energy as function of displacements
even if the band structure is somehow distorted in the
Hartree-Fock calculation (see Ref.~\onlinecite{indo} for the
discussion to this point for KNbO$_3$). In tuning the INDO
parameters, we primarily use earlier FP-LMTO calculations as a benchmark,
also with some reference to experimental data (e.g., for phonons
in KTaO$_3$).

The objectives of the present study are the total energy as a function
of off-center Li displacements in KTaO$_3$, the 90$^0$-energy barrier
in the hopping motion of displaced Li, and the lattice relaxation
around the Li impurity.
As a by-product, we provide the description of pure KTaO$_3$
by the INDO method, including the calculation of $\Gamma$ phonons.
The paper is organized as follows. In Sec.~\ref{sec:param},
we discuss the choice of INDO parameters for our calculations
of pure and Li-doped KTaO$_3$. In Sec.~\ref{sec:phonon},
we present the calculated $\Gamma$ transversal optic (TO)
phonon frequencies and eigenvectors in KTaO$_3$.
In Sec.~\ref{sec:klt}, we discuss the off-center Li displacement
and the lattice relaxation around a Li impurity in KTaO$_3$.

\begin{table}[b]
\caption{One-center INDO parameters for Ta and Li}
\begin{tabular}{lcdddd}
 Orbital & \mbox{$\zeta$ (a.u.$^{-1}$)} & $E_{\text{neg}}$ (eV) &
	   $-\beta$ (eV) & \mbox{$P_0$ (a.u.)} \\
\hline
  Ta~$6s$ &  2.05  &     0.0  &  20.0  &  0.1   \\
  Ta~$6p$ &  2.05  &  $-$2.0  &  20.0  &  0.0   \\
  Ta~$5d$ &  1.70  &    23.50 &  16.0  &  0.6   \\
  Li~$2s$ &  1.10  &     5.0  &   0.6  &  0.1   \\
  Li~$2p$ &  0.80  &     0.9  &   0.6  &  0.02  \\
\end{tabular}
\label{tab:param}
\end{table}

\section{INDO parameterization for KT\lowercase{a}O$_3$}
\label{sec:param}

The description of the INDO approximation to the
Hartree-Fock--Roothaan method is given in Refs.~\onlinecite{indo1,indo2}.
The major formulae defining the parameters of the calculation
are also cited in Ref.~\onlinecite{indo}, where the choice of parameters
relevant for the calculation of KNbO$_3$ is discussed.
As in Ref.~\onlinecite{indo}, we performed the present
calculations with the CLUSTERD computer code
by Shidlovskaya, Shluger and Stefanovich\cite{indo1,indo2}.
The parameters obtained there for K and O atoms are retained in the
present calculation. In the following, we refer to
Ref.~\onlinecite{indo} for the description of relevant parameters.
For KTaO$_3$, we had to specify $\zeta_{\mu}$ (parameter of
the Slater exponent), $E_{neg}$ (central energy position),
$\beta_{\mu}$ (parameter of the resonance interaction with
other states) and $P^{(0)}$ (occupation number) for the Ta and Li
states. Moreover, the two-center parameter $\alpha_{AB}$ had
to be specified for Ta and Li in combinations with other atoms.
The INDO calculations for Li halides\cite{indo2,lif} and
Li$_2$SiO$_3$~\cite{indo1} have been done earlier, and some
parameters tabulated. In contrast to these earlier works, we
preferred to use both Li~$2s$ and $2p$ states in the basis set,
since the bonding in KTaO$_3$:Li is much less ionic than in
Li halides. Moreover, a very delicate balance of the interactions
resulting in a ferroelectric instability favors the use of
a more extended basis set.
The recommended values of one-center Li parameters which are
provided in the description of the INDO code~\cite{code}
were found to be a reasonable choice (see below),
and some discussion as for the choice of two-center
$\alpha_{\mbox{\small O--Li}}$ parameter is given in Sec.~\ref{sec:klt}.
As for the Ta-related parameters, we took $E_{neg}$ and $P^{0}$ values
(with the exception of $E_{neg}$ for Ta~$5d$) the same as for
corresponding ($5s$, $5p$ and $4d$) Nb states in KNbO$_3$,
based on the similarity of the band structures.
({\it Ab initio} calculations of the band structure of both
compounds have been performed earlier, e.g.,
by Neumann {\it et al.}\cite{asa92}).

\begin{table*}[t]
\caption{
Calculated $\Gamma$-TO frequencies and eigenvectors
of the $T_{1u}$ modes in cubic KTaO$_3$.
}
\label{tab:phonon}
\begin{tabular}{ddddcc}
\multicolumn{4}{c}{Eigenvectors} \\
\cline{1-4}
  K & Ta & O$_1$ & O$_{2,3}$ &
\raisebox{2.5ex}[0pt]{$\omega$ calc. (cm$^{-1}$)} &
\raisebox{2.5ex}[0pt]{$\omega$ exp. (cm$^{-1}$)} \\
\hline
 0.06 & $-$0.46 & 0.71 & 0.37 & 86$^a$ \\
 0.11 & $-$0.50 & 0.47 & 0.51 & 80$^b$ &
\raisebox{2.5ex}[0pt]{25--106$^c$; 81$^d$; 85$^e$} \\
\hline
 $-$0.92 & 0.31 & 0.14 & 0.13 & 202$^a$ \\
 $-$0.91 & 0.28 & 0.14 & 0.17 & 172$^b$ &
\raisebox{2.5ex}[0pt]{196--199$^c$; 199$^d$; 198$^e$} \\
\hline
 $-$0.01 & 0.13 & $-$0.65 & 0.53 & 500$^a$ \\
    0.01 & 0.09 & $-$0.83 & 0.39 & 528$^b$ &
\raisebox{2.5ex}[0pt]{551--550$^c$; 546$^d$; 556$^e$} \\
\end{tabular}
$^a$ Present work. \\
$^b$ Full-potential LAPW calculation of Ref.~\onlinecite{singh}. \\
$^c$ Infrared reflectivity measurements at 12$-$463 K,
     Ref.~\onlinecite{pmn67}. \\
$^d$ Hyper-Raman scattering measurements at room temperature,
     Ref.~\onlinecite{vu84}. \\
$^e$ Raman scattering measurements at room temperature (soft mode)
     and at 10 K, Ref.~\onlinecite{fw68}.
\end{table*}

The final choice of the INDO parameters for KTaO$_3$ was done based
on a calculation of frozen $\Gamma$-TO phonons. Since the three
phonon modes which belong to the $T_{1u}$ irreducible representation
are different combinations of the vibrations of K, Ta, and O,
they provide a good indication of the accuracy with which
all essential interatomic interactions are treated.
The TO phonon frequencies are measured in a number of experiments,
and their eigenvectors determined in {\it ab~initio} calculations.
This makes the adjustment of $\alpha_{AB}$ and $\beta_{\mu}$
parameters, which affect the shape of the total energy hypersurface
as function of a particular displacement and changes the frequencies
and eigenvectors, a relatively easy task. To begin with, we
took as a benchmark the total energy difference as function of individual
displacements, determined in the full-potential LMTO
calculation.\cite{ktn3} However, finally we aimed at obtaining the phonon
eigenvectors with apparently even better accuracy, as calculated
in Ref.~\onlinecite{singh} by the full-potential linear
augmented plane waves (LAPW) method.
The values of the \mbox{INDO} parameters we found to be best suited
for the study of KTaO$_3$:Li (in addition to those published
already in Ref.~\onlinecite{indo}) are listed in Table~\ref{tab:param}.
The two-center parameters $\alpha_{AB}$ which account for the non-point
character of the interaction of a valence orbital at the atom $A$
with the core of atom $B$ (see explicit definition in
Refs.~\onlinecite{indo2,indo}) are
0.023, 0.15, 0.39 and 0.58 for $A$=O and
$B$=Li, O, K and Ta, correspondingly, and zero for $A$=Ta, K, or Li.

The band gap in KTaO$_3$ as calculated by the INDO method
is 6.7 eV, that is close to, but larger than, 6.1 eV as obtained
in a similar calculation for KNbO$_3$\cite{indo}.
The absolute value of the gap in the one-electron
energy spectrum is known to come out systematically larger
in the Hartree-Fock formalism as compared with spectroscopic data,
because the unscreened Coulomb interaction
shifts the unoccupied states too high in energy.
Nevertheless, the difference between the gap values for two compounds
is in agreement both with the experimental estimates
(3.3~eV for KNbO$_3$ vs. 3.8~eV for KTaO$_3$) and the results of
the calculations done in the local density approximation
(1.4~eV vs. 2.1~eV\cite{asa92,singh}).
The (static) effective charges for KTaO$_3$, as estimated
from the INDO calculations based on the Mullikan population
analyzis, are $+$0.62 (K), $+$2.23 (Ta) and $-$0.95(O).
This reveals a somehow increased ionicity for KTaO$_3$,
in comparison with KNbO$_3$\cite{indo}, in general agreement
with the tendency pointed out by Singh\cite{singh}.

\section{$\Gamma$-TO phonons in KT\lowercase{a}O$_3$}
\label{sec:phonon}

$\Gamma$-TO phonons in the cubic perovskite structure are split
by symmetry into three triple degenerate $T_{1u}$ modes and
one triple degenerate $T_{2u}$ mode. {\it Ab initio} calculations
of frequencies and eigenvectors have been done by Postnikov
{\it et al.} \cite{phonon} and by Singh \cite{singh}.
Whereas calculated frequencies
are in reasonable agreement between both calculations
and with the experimental data available, there is some
disagreement in the estimations of calculated eigenvectors,
especially for the soft mode, as was discussed at length
in Ref.~\onlinecite{singh}. The full-potential LAPW
method is able to provide ultimately better accuracy
than LMTO, due to a more extended basis set.
Moreover, the indirect experimental indications
of the displacements within the soft mode of KTaO$_3$\cite{per89}
seem to be in agreement with the LAPW results. Therefore we aimed
at reproducing the eigenvectors of Ref.~\onlinecite{singh}
with the proper choice of our INDO parameters.
We performed our frozen phonon calculations at the
experimental lattice constant (extrapolated to zero temperature)
of 3.983\AA. The elements of the dynamical matrix were found
from the polynomial fit of the total energy as function
of various displacements within the $T_{1u}$ mode. Calculated
frequencies and eigenvectors are shown in Table~\ref{tab:phonon}.
O$_1$ stands for the oxygen atom at $(\frac{1}{2}\frac{1}{2}0)$,
and O$_{2,3}$ for two (equivalent) atoms at
$(0\frac{1}{2}\frac{1}{2})$ and $(\frac{1}{2}0\frac{1}{2})$,
for the vibrations along $[001]$. K is at $(000)$
and Ta at $(\frac{1}{2}\frac{1}{2}\frac{1}{2})$ of the cubic
perovskite cell.

The agreement with the eigenvectors of Singh\cite{singh}
is good, especially in what regards the relative displacement
of potassium and tantalum with respect to each other and to
the averaged displacement of the oxygen sublattice.
The difference in the displacement patterns of O$_1$
and O$_{2,3}$ atoms within the soft mode seems, however,
to be slightly overestimated in our case.

For the $T_{2u}$ mode, our calculation gives the frequency of
260 cm$^{-1}$, as compared to 264 cm$^{-1}$ by Singh\cite{singh}
and the experimental estimations
of 264 cm$^{-1}$ (Ref. \onlinecite{per89})
to 274 cm$^{-1}$ (Ref.~\onlinecite{vu84}). This mode involves
only the stretching within the O$_{2,3}$ sublattice, therefore
the eigenvector is uniquely defined. The excellent agreement of
the frequency with the experimental value(s) suggests that
our choice of the INDO parameters, that describe the O -- O interaction
as mediated by the central Ta atom, is sufficiently good.

\section{L\lowercase{i} off-center displacement and lattice relaxation}
\label{sec:klt}

With the parameters of the INDO method properly tuned,
one can obtain better accuracy in describing energy
trends and lattice relaxation in Li-doped KTaO$_3$ than was possible
with the simple polarizable point-charge model\cite{vdkk84}
or within the shell model\cite{stac90,ex:Li,stac91}).
On the other hand, the INDO calculation allows to overcome
the problems of previous first-principles FP-LMTO
studies\cite{wil94} in what regards small supercell size
and the problems of the {\it muffin-tin}-spheres packing.
Since the {\it muffin-tin} geometry is not used in a tight-binding
INDO scheme, the problems of spheres packing do not exist there;
at the same time, the method
is much less computationally demanding and allows to treat
larger supercells (all results discussed below refer
to the $3\!\times\!3\!\times\!3$ KTaO$_3$ supercell,
i.e. that with 135 atoms in total, with one substitutional
Li impurity). It is also possible to search for the optimized
lattice distortion around the displaced impurity in the course
of INDO calculation, that was not done in Ref.~\onlinecite{wil94}.

\begin{figure}
\epsfxsize=8.5cm
\centerline{\epsfbox{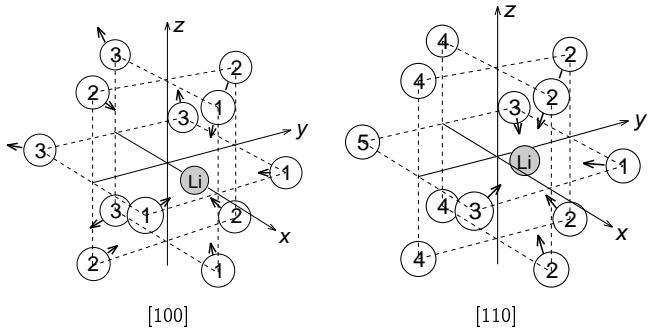}}
\vspace*{0.5cm}
\caption{[100] and [110] Li off-center displacements
and the relaxation pattern of neighboring oxygen atoms.
}
\label{fig:struc}
\end{figure}

\begin{figure}
\epsfxsize=8.2cm
\centerline{\epsfbox{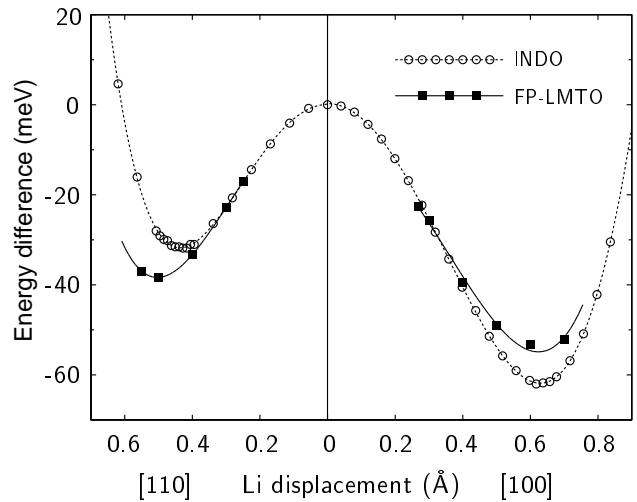}}
\vspace*{0.5cm}
\caption{Total energy gain as function of [100] and [110] Li displacements
calculated by FP-LMTO method
(Ref.~\protect{\onlinecite{wil94}})
and by INDO for the $3\!\times\!3\!\times\!3$ supercell.
The curves are shifted so as to match for the 0.3~{\AA} displacement.
}
\label{fig:compar}
\end{figure}

In addition to the parameters which relate to the KTaO$_3$ bulk
and which were tuned on the basis of frozen phonon calculations
of Sec.~\ref{sec:param}, and the one-site Li parameters
borrowed from earlier INDO calculations for Li-containing
ionic crystals\cite{indo1,indo2}, we had to specify the two-center
parameter $\alpha_{\mbox{\small O--Li}}$, that enters the empirical expression
for the interaction of the O valence electrons with the Li
core and that effectively accounts for the finite size
of the Li core and for the diffuseness of the O-related
valence band states (see definition of this parameter and
the discussion in Ref.~\onlinecite{indo2}).
Whereas already the first choice of $\alpha_{\mbox{\small O--Li}}$=0
as proposed in Ref.~\onlinecite{indo2} provides a qualitatively correct
effect on the equilibrium geometry (Li displaces off-center
along [100] by $\sim$0.4 {\AA}), we found that a fine tuning
of this parameter has a considerable effect on the
magnitude of the displacement, and especially on the energy
gain associated to it. For instance, $\alpha_{\mbox{\small O--Li}}$=0.02
sets the energy gain at $\sim$60 meV at the $\sim$0.6~{\AA}
Li displacement, whereas $\alpha_{\mbox{\small O--Li}}$=0.04
lowers the total energy by $\sim$100~meV at the $\sim$0.7~{\AA} displacement.
The best fit to the results of the FP-LMTO calculation\cite{wil94}
(that were done only for the Li displacement $\ge$0.3~{\AA})
occurs at $\alpha_{\mbox{\small O--Li}}$=0.023.
We attempted to reproduce the FP-LMTO results for [100]
and [110] Li off-center displacements -- the former being related
to the true ground-state configuration, and the latter to
the saddle point between two adjacent displaced Li positions.
The displaced Li ion with its neighboring oxygen atoms
is shown schematically in Fig.~\ref{fig:struc}.
The total energy as function of [100] and [110] Li off-center
displacements from FP-LMTO and from our present calculations
is shown in Fig.~\ref{fig:compar}.

Another sensitive question is the dependence of the results
on the supercell size. The INDO method in the implementation
we use produces in each iteration the one-electron energy spectrum
in the $\Gamma$ point only. The calculation is normally
being done with an extended supercell, so that the band dispersion
over the correspondingly shrinked Brillouin zone is neglected.
The enlargement of a supercell in the calculation for a
perfect crystal is in a sense analogous to choosing a finer mesh
for the {\bf k}-space integration in a conventional band structure
calculation. When treating an impurity system, another aspect
becomes essential as well, that is, the effect of interaction
between impurities situated in adjacent supercells should be
kept negligible, or at least controllable. For the KTaO$_3$:Li system,
it means that the polarization cloud associated with a single
off-center displaced Li impurity should be, in the ideal case,
fully within the supercell chosen.
The size of the polarized region associated with the [100]-displaced
Li ion was estimated in the shell model calculation by Stachiotti
and Migoni\cite{stac90} to be about 5 lattice constants
along the direction of displacement, with $\sim$99\% of the
`effective dipole' polarization being confined to nearest
Ta--O chains, that go parallel to the displacement.
As is discussed below, the magnitudes of the atomic displacements
and polarization in our present calculation is considerably
smaller than those found in Ref.~\onlinecite{stac90}, and
the relaxed neighbors to the Li impurity are well within
the $3\!\times\!3\!\times\!3$ supercell. In order to be on safer side,
we performed as well the calculations for a supercell doubled
in the direction of Li displacement, i.e., $6\!\times\!3\!\times\!3$,
with a single [100]-displaced Li atom.
The equilibrium displacement in this case is 0.62~{\AA},
exactly as for the $3\!\times\!3\!\times\!3$ supercell
(see Fig.~\ref{fig:compar}),
with the energy lowering 57.2~meV. The difference
from the result for a $3\!\times\!3\!\times\!3$ supercell (62.0~meV)
roughly represents the uncertainty related to the supercell size
in our calculations.

The magnitude of the Li off-center displacement naturally
agrees well with the FP-LMTO data (0.61~{\AA}\cite{wil94}),
since the latter was an important benchmark in our choice
of INDO parameters.
We failed however to obtain the ideal matching with the FP-LMTO
results in what regards both the off-center displacement {\it and}
the energy gain in {\it both} [100] and [110] directions, as is seen
in Fig.~\ref{fig:compar}.
Our equilibrium [100] off-center displacement
is smaller than the estimate (1.35~{\AA})
of the polarizable point-charge model\cite{vdkk84},
or 1.44~{\AA} as calculated by Stachiotti and Migoni
within the shell model\cite{stac90}.
On the other hand, our value is in good agreement with
a more recent, and apparently more elaborately parameterized, shell model
calculation by Exner {\it et al.} (0.64~{\AA}, Ref.~\onlinecite{ex:Li}).

The energy gain due to the Li off-center displacement
is not directly measurable in an experiment, but there are
estimations for the $90^0$-energy barrier
between, say, [100] and [010]-displaced positions
to be 86~meV\cite{vdkk84}. The hoppings between such adjacent positions
may only occur via the [110] saddle point (see Fig.~\ref{fig:compar}).
Our estimate of the energy difference between [100] and [110] minima
is $\sim$30.2~meV, roughly two times larger than in the FP-LMTO
calculation\cite{wil94}, but much less than the experimental
estimate. The origin of this discrepancy, as has been mentioned
in Ref.~\onlinecite{wil94}, is most probably related to the lattice relaxation
around the displaced Li ion, that makes the net energy gain
from the displacement larger, and the $90^0$-activation energy
(involving now the displacement of many atoms) correspondingly higher.
Indeed, the second harmonic generation-based estimates of the activation
barrier\cite{voigt} reveal two types of processes, apparently
one involving the lattice relaxation (with the barrier height 86.2~meV)
and another one that is too fast for the lattice to follow,
with the barrier 14.7~meV.

\begin{figure}
\epsfxsize=8.2cm
\centerline{\epsfbox{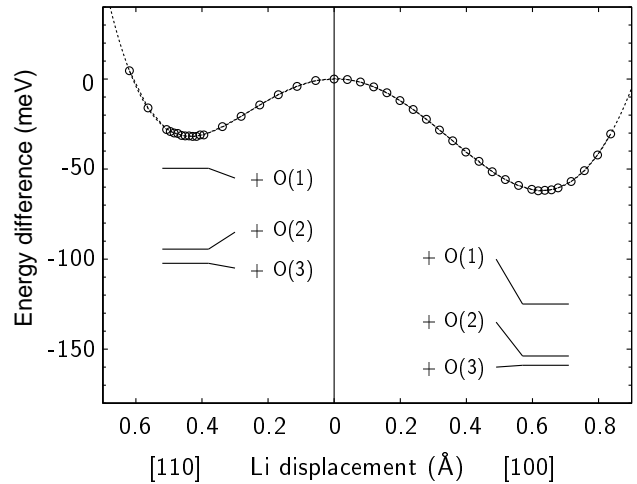}}
\vspace*{0.5cm}
\caption{Total energy gain as function of [100] and [110] Li displacements
without lattice relaxation (dashed line with open circles),
and the total energy values after including the relaxation
of three groups of nearest oxygen atoms.
}
\label{fig:relax}
\end{figure}

In order to clarify this point, we performed a lattice relaxation
of several shells of neighbors to the displaced Li ion, for
the cases of [100] and [110] displacements. The relaxed
coordinates of atoms are given in Tables \ref{tab:relax100}
and \ref{tab:relax110}, where the oxygen atoms are numbered
consistently with Fig.~\ref{fig:struc}.
The total energy values resulting
from the gradual inclusion of neighbor relaxation are shown
in Fig.~\ref{fig:relax}. We found the relaxation of twelve
nearest oxygen atoms essential, and the effect of relaxing nearest Ta
and more distant atoms to be negligible, in what regards the
effect on the total energy.
The energy gain in the fully relaxed [100]-displaced configuration,
with respect to a non-relaxed central Li position, is 158.9~meV;
the energy gain in the relaxed [110]-configuration is 102.3~meV.
Therefore, the enhancement of the excitation barrier due to
relaxation effects is by a factor of two, but still not
sufficient to reach experimentally expected $\sim$86 meV.
This discrepancy may be due to the fact that in reality
the $90^0$-reorientation process of the impurity does not
necessarily occur via the fully relaxed saddle-point configuration.
Depending on the actual degree of relaxation around the saddle-point
Li position, the barrier height is expected from Fig.~\ref{fig:relax}
to be between $\sim$57 meV (full relaxation at the saddle point)
to $\sim$127 meV (no relaxation).

\begin{table*}[t]
\caption{
Relaxed atomic positions for the [100] Li displacement
as calculated by INDO.
}
\label{tab:relax100}
\begin{tabular}{ccccc}
Atom & \multicolumn{3}{c}{Lattice coordinates} & Displacement \\
\hline
           Li  & $\Delta_x$ & 0 & 0 & $\Delta_x=0.1550$ \\
 4$\times$O(1) & $\frac{1}{2}+\Delta_x$ & $\frac{1}{2}+\Delta_y$ & 0 &
 $\Delta_x=-0.0045$; $\Delta_y=-0.0105$ \\
 4$\times$O(2) & $\Delta_x$ & $\frac{1}{2}+\Delta_y$ &
			       $\frac{1}{2}+\Delta_y$ &
 $\Delta_x=0.0070$; $\Delta_y=-0.0026$  \\
 4$\times$O(3) & $-\frac{1}{2}+\Delta_x$ & $\frac{1}{2}+\Delta_y$ & 0 &
 $\Delta_x=-0.0020$; $\Delta_y= 0.0020$ \\
\end{tabular}
\end{table*}

\begin{table*}[t]
\caption{
Relaxed atomic positions for the [110] Li displacement
as calculated by INDO.
}
\label{tab:relax110}
\begin{tabular}{ccccc}
Atom & \multicolumn{3}{c}{Lattice coordinates} & Displacement \\
\hline
           Li  & $\Delta_x$ & $\Delta_x$ & 0 & $\Delta_x=0.0760$ \\
 1$\times$O(1) & $\frac{1}{2}+\Delta_x$ & $\frac{1}{2}+\Delta_x$ & 0 &
 $\Delta_x=-0.0090$ \\
 4$\times$O(2) & $\frac{1}{2}+\Delta_x$ & $\Delta_y$ &
			       $\frac{1}{2}+\Delta_z$ &
 $\Delta_x=-0.0060$; $\Delta_y=0.0030$; $\Delta_z=-0.0080$  \\
 2$\times$O(3) & $\frac{1}{2}+\Delta_x$ & $-\frac{1}{2}+\Delta_y$ & 0 &
 $\Delta_x=-0.0020$; $\Delta_y=0.0060$ \\

 4$\times$O(4) & $-\frac{1}{2}+\Delta_x$ & $\Delta_y$ &
			       $-\frac{1}{2}+\Delta_z$ &
 $\Delta_x=0.0003, \Delta_y=0.0001, \Delta_z=0.0003$  \\

 1$\times$O(5) & $-\frac{1}{2}+\Delta_x$ & $-\frac{1}{2}+\Delta_x$ & 0 &
 $\Delta_x \sim 0$ \\
\end{tabular}
\end{table*}

It should be noted that the spatial range
of polarization is not so large in our calculation
as follows from the shell model results of both Ref.~\onlinecite{stac90}
and \onlinecite{ex:Li}, and the displacements of twelve nearest
oxygen atoms around the Li ion are, generally, smaller in our case.
There are also some qualitative differences in the displacement pattern.
The largest striking, the O(2) atoms (as labeled in
Table~\ref{tab:relax100}) follow the Li displacement in our
calculation, whereas they move away from the displaced Li ion
in both shell model calculations cited. This is probably
due to an oversimplified parameterization of the interaction
potential used in the shell model. In the INDO method,
no special approximation is introduced for the description
of the chemical bonding, therefore our results seem
to be more reliable. The numerical values of the displacements
may, however, be somehow refined in subsequent calculations
with larger unit cells.

\section{Summary}

As an extension of our previous study of ferroelectric KNbO$_3$
with a semiempirical INDO method, we performed calculations
for pure paraelectric KTaO$_3$, concentrating on TO phonon
frequencies as a benchmark for fine tuning of our INDO
parameterization. In a series of supercell calculations
for Li-doped KTaO$_3$, tuned in such a way as to reproduce
the energetics of the Li off-center displacement previously
found in FP-LMTO calculations, we analyze the relaxation
of near neighbors to the Li impurity, and the impact of this
relaxation on the reorientational energy barriers.
The relaxation pattern in some aspects differs from that calculated
earlier within the shell model.
The study of the interaction between Li impurities
in a polarized lattice seems feasible with the method used.

\acknowledgements
The work has been done as part of the German-Israeli
joint project ``Perovskite-based solid solutions
and their properties''. Financial support by the
Nieders\"achsische Ministerium f\"ur Wissenschaft und Kultur
and by the Deutsche Forschungsgemeinschaft
(SFB~225) is gratefully acknowledged. The authors are
grateful to E.~Stefanovich for fruitful discussions.

\end{document}